\documentclass[amsmath,amssymb,aps,onecolumn,prr,freq,floatfix]{revtex4-2}
\usepackage{graphicx}% Include figure files
\usepackage{dcolumn}% Align table columns on decimal point
\usepackage{bm}% bold math
\usepackage{hyperref}% add hypertext capabilities
\usepackage{color}
\usepackage{subfigure}
\usepackage{flexisym}
\usepackage{physics}

\begin{document}

\title{Precessing magnetic particles as ac magnetic field sensors}

\author{A. T. M. Anishur Rahman}
\altaffiliation{Department of Physics, University of Warwick, Coventry, UK}
\email{anishurrahman91@gmail.com}
%Lines break automatically or can be forced with 
%
%  \email{Second.Author@institution.edu}

%%ly specified

\begin{abstract}
Electromagnetic waves are widely used including in defense, biomedicine, and fundamental science. Their efficient detection determines how we communicate, defend against adversaries, diagnose diseases and perform search and rescue operations. In this article, exploiting the precession of a levitated magnetic particle in vacuum, we show that weak electromagnetic waves down to the femtotesla level can be detected. It is also shown that such a sensor has a large dynamic range over a millitesla, is continuously tunable over many gigahertz and can detect frequencies with sub-hertz resolutions. The direction of arrival of the incoming electromagnetic wave can also be found relatively easily.
\end{abstract}
\maketitle

%%Usage of electromagnetic waves is ubiquitous and 
%%Electromagnetic (EM)  waves are widely used for, among others, wireless communications \cite{Tse2005}, radar \cite{Spezio2002}, biomedicine \cite{LiMW2013, Zhao2022}, and search and rescue \cite{NguyenMW2019}.
\section{Introduction}
The ability to detect electromagnetic (EM) fields has applications, among others, in fundamental physics \cite{Penzias1965,Thornton2013,AlesiniPRD2021,RahmanPRD2022}, radar \cite{Spezio2002}, biomedicine \cite{LiMW2013, Zhao2022,GORYANIN2020757}, search and rescue \cite{NguyenMW2019}, climate monitoring \cite{Chelton2005} and wireless communications \cite{Tse2005}. In fundamental physics, microwave sensors have been used for the detection of cosmic microwave background \cite{Penzias1965} and fast radio bursts \cite{Thornton2013} and suggested for the search for dark matter \cite{AlesiniPRD2021,RahmanPRD2022,SikivieRMP2021}. In biomedicine, EM sensors are used for monitoring various aspects of our health, including cardiopulmonary activities \cite{LiMW2013, Zhao2022} and breast cancer \cite{GORYANIN2020757}, whilst in search and rescue operations, such sensors are deployed to detect living things hidden beneath rubble \cite{NguyenMW2019}. Likewise, the detection of civil and military aircraft is routinely performed using EM sensors \cite{Spezio2002}. Traditional EM sensors include antennas that, once made, cannot be changed and have a limited frequency range of operation. 

%The detection of electromagnetic (EM) waves is essential for, among others, fundamental science \cite{Penzias1965,Thornton2013,AlesiniPRD2021,RahmanPRD2022}, wireless communications \cite{Pedersen2023}, radar \cite{Spezio2002}, biomedicine \cite{LiMW2013, Zhao2022} and search and rescue \cite{NguyenMW2019}. 

In principle, EM fields can be detected using either an electric or a magnetic field sensor. Examples of electric field sensors include Rydberg atoms \cite{FancherIEE2021, JingMingyong2020Asrb,GordonAIPAdv2019}. In Autler--Townes configuration, such a sensor can detect discrete frequencies between MHz and THz~\cite{Yuan_2023,FancherIEE2021} and fields as weak as $\approx$5~$\mu$V/m ($1.7\times 10^{-14}~$T) \cite{GordonAIPAdv2019,JingMingyong2020Asrb,FancherIEE2021}. Existing magnetic field sensors, such as atomic vapors \cite{BudkerDmitry2007Om} and the nitrogen-vacancy (NV) center in diamond-based sensors~\cite{Graham2023} are predominantly used as DC field sensors, although some progress has been made towards the detection of EM fields using NV centers in diamond \cite{Meinel2021}. Such magnetic field sensors have a relatively small dynamic range, e.g., $\le$$\mu$T. Superconducting quantum interference devices are excellent magnetic field sensors but require cryogenic temperatures \cite{Couedo2019}. Efficient sensing of EM waves can enhance our capability to defend against adversaries by detecting the weakest possible signals and hence providing early warnings \cite{Spezio2002}, perform better search and rescue operations by detecting faint signals from living things hidden under rubble \cite{NguyenMW2019}, and diagnose diseases \cite{GORYANIN2020757,LiMW2013, Zhao2022}. Re-configurable detectors, in particular detectors that can be configured for different frequencies on demand while in operation, would also benefit the aforementioned areas.

Levitation in a vacuum provides a contactless and near-frictionless environment. This makes levitated particles susceptible to external stimuli, making them extremely good sensors. For example, using the center-of-mass motions of such particles, zeptonewton scale force sensitivity has been achieved \cite{RanjitPRA2016}. Likewise, exploiting the rotational motion of a levitated particle, an extremely small torque has been measured \cite{AhnNatNano2020}. Among levitated particles, magnetic particles are unique in the sense that they contain an extra degree of freedom, i.e., the spin, which makes them even more versatile. The coupling between the spin and the other degrees of freedom of a levitated magnetic particle has not been explored yet, but is promising for developing new technologies \cite{JacksonKimballPRL2016} and exploring fundamental physics~\cite{Rahman_2019}.

%%Recently, levitated magnets in cryogenic conditions have been used for detecting DC magnetic fields \cite{Ahrens2024levitated}.

In this article, using the precessional motion of a levitated magnetic particle in high vacuum, it is shown that extremely weak electromagnetic waves of femtotesla strength can be detected. Such a magnetometer has a dynamic range over a millitesla, can be continuously tuned over many GHz, and detects frequencies with sub-hertz resolutions. It is also shown that the direction of arrival of the EM wave can be determined relatively easily.

\begin{figure}
%    \centering
    \includegraphics[width=8.5cm]{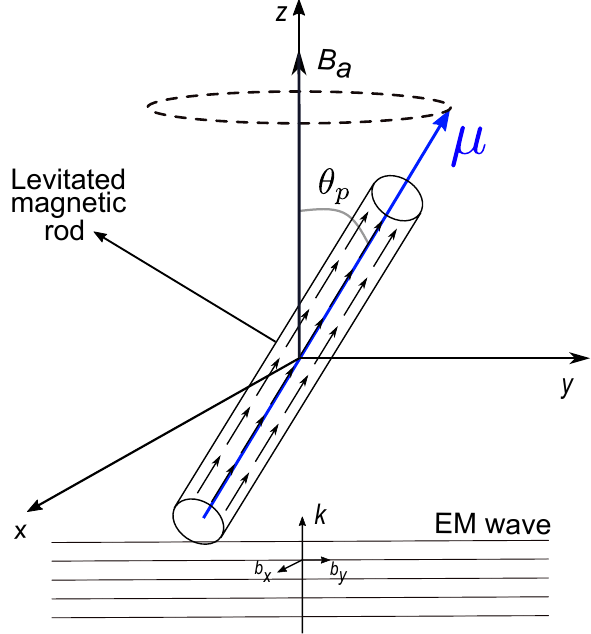}
    \caption{Precession %MDPI: Please note: 1. Figures should be cited after where they are first mentioned, we have adjusted all. 2. The size and position of figures will be adjusted appropriately to avoid blank space and to ensure the clarity and readability of the image during the final steps. 3. Please check all figures carefully, including the content in the figure, subfigures explanation, number, resolution, etc. After being confirmed and published, we cannot update the figures.
 of a levitated magnet when exposed to circularly polarized electromagnetic waves. A homogeneous DC magnetic field $B_a\hat{z}$ is applied along the $z~$axis. A circularly polarized electromagnetic (EM) wave propagating along the $z~$axis initiates spin precession, which subsequently induces a mechanical precession. The mechanical angle of precession is denoted by $\theta_p$. The wavevector of the electromagnetic wave is represented by $\mathbf{k}$. The magnetic field associated with the EM wave is in the $xy~$plane with $b_x=b_0\cos\omega t$, $b_y=\sigma b_0\sin\omega t$ and $\sigma=\pm 1$. When $\sigma=+1$ ($\sigma=-1$), the incoming EM wave rotates anticlockwise (clockwise).} 
    \label{fig1}
\end{figure}

\section{Theoretical Model}
Consider a cylindrical rod (see Fig. \ref{fig1}) of mass $M$, mass density $\rho_m$, radius $r$, length $L$ and magnetic moment $\boldsymbol{\mu}$ polarized along its easy magnetization axis, which is levitated in ultra high vacuum (UHV). A homogeneous DC magnetic field $\mathbf{B_a}$ (Figure \ref{fig1}) is applied along the $+z~$axis, which ensures $\boldsymbol{\mu}$ aligns with $B_a$, i.e., $\boldsymbol{\mu}=[0~0~\mu_s]$, where $\mu_s$ is the saturated magnetic moment of the levitated particle. Consider also that a circularly polarized electromagnetic wave $\mathbf{b_s}=b_0 \Re{(\exp(i\omega t))}~\hat {x} +\sigma b_0\Re{(i\exp(i\omega t))}~\hat {y}$ whose frequency ($\omega$) and strength ($b_0$) we aim to determine is propagating in the $+z$ direction and illuminates the levitated magnetic particle where $\sigma =-1 ~(+1)$ corresponds to clockwise (anticlockwise) rotation. It is expected that $b_s\ll B_a$. The interaction between the EM wave and the spins or magnetization initiates a spin precession in the ferromagnetic particle \cite{KittelSolidState,Gilbert2004} giving rise to components of magnetic moment in the $xy~$plane, i.e., $\mu_x$ and $\mu_y$. In a ferromagnetic material, spins and thus magnetization are connected to the crystal lattice via magnetocrystalline anisotropy \cite{KittelSolidState}. This provides a link between the internal (spin) and the mechanical degrees of freedom of a magnetic object. When the moment of inertia $I$ of the levitated object and the rotational damping $\Gamma_g$ due to gas molecules that it encounters are sufficiently low, the levitated object starts to precess with the magnetization \cite{KeshtgarPRB2017,JacksonKimballPRL2016,Fadeev_2021}. The precession of the levitated object, however, induces magnetization through the Barnett effect and perturbs the overall spin dynamics \cite{KeshtgarPRB2017}. For the analysis that follows, we do not consider the translational motions of levitated particles and the fluctuating torque arising from gas molecules, as such torques in UHV are relatively small (see Section \ref{sec3} and Equation~(\ref{eqn8}) for further discussion). The dynamics of the spin and mechanical precession can be modeled using the Landau--Lifshitz--Gilbert equation and the rotational equation of motion \cite{Gilbert2004,KeshtgarPRB2017,LyutyyPRB2019,KaniTwamleyPRL2022}, respectively, and are written as

%MDPI: Sub Equations are not recommended. Please try to renumber all equations in numerical order.
%MDPI:  1. Please double check and unify the format of all variables carefully (italics, bold, superscript, subscript, uppercase, lowercase, etc.) 2. Please ensure there are no duplicate equations.

% \begin{subequations}
% \begin{align}
%     \frac{1}{\gamma}\frac{d\boldsymbol{\mu}}{d t}&=\boldsymbol{\mu}\times \mathbf{B_e}-\frac{\alpha}{\gamma\mu_s}\Bigl[\boldsymbol{\mu}\times \Bigl[\frac{d\boldsymbol{\mu}}{d t}-\boldsymbol{\Omega}\times \boldsymbol{\mu}\Bigr]\Bigr],
%     \label{eqn0}\\
%     I \dot{\boldsymbol{\Omega}}&= \frac{1}{\gamma}\frac{d\boldsymbol{\mu}}{d t}+\boldsymbol{\mu} \times \mathbf{B}-I\Gamma_g\boldsymbol{\Omega},
%     \label{eqn00}
%     \end{align}
% \end{subequations}

\begin{eqnarray}
   \frac{1}{\gamma}\frac{d\boldsymbol{\mu}}{d t}&=\boldsymbol{\mu}\times \mathbf{B_e}-\frac{\alpha}{\gamma\mu_s}\Bigl[\boldsymbol{\mu}\times [\frac{d\boldsymbol{\mu}}{d t}-\boldsymbol{\Omega}\times \boldsymbol{\mu}]\Bigr],
    \label{eqn0} 
\end{eqnarray}

\begin{eqnarray}
    I \frac{d\boldsymbol{\Omega}}{dt}&= \frac{1}{\gamma}\frac{d\boldsymbol{\mu}}{d t}+\boldsymbol{\mu} \times \mathbf{B}-I\Gamma_g\boldsymbol{\Omega},
    \label{eqn00}
   \end{eqnarray}

where $\gamma$ is the gyromagnetic ratio, $\boldsymbol{\Omega}$ is the angular frequency vector, $\mathbf{B_e}=\mathbf{b_s}+(B_a+B_{an})~\hat{z}$, $\mathbf{B}=\mathbf{b_s}+\mathbf{B_a}$ with $B_{an}$ being the field associated with the magnetocrystalline anisotropy~\cite{BryantAPL1992}, and $\alpha > 0$ is the dimensionless Gilbert damping constant. Since $b_0\ll B_a$ we make the simplifying assumptions $\mu_z\approx \mu_s$ and $\dot{m_z}=0$, $\Omega_z=0$ and $\dot{\Omega_z}=0$ \cite{KeshtgarPRB2017} and solve the coupled differential Equations (\ref{eqn0}) and (\ref{eqn00}) using the ansatz $\mu_x=\mu_0\exp(i\omega t)$, $\mu_y=i\mu_0\exp(i\omega t)$, $\Omega_x=\Omega_0 \exp(i\omega t)$ and $\Omega_y=i\Omega_0 \exp(i\omega t)$. We find 
\begin{eqnarray}
\mu_0&=&\frac{b_0\gamma \mu_s\sqrt{1+(\frac{\Gamma_g}{\omega}+\frac{\alpha \mu_s}{\gamma\omega I})^2}~~\exp{(i\theta_m)}}{\sqrt{\bigl((B_a+B_{an})\gamma+\alpha\Gamma_g +\sigma \omega\bigr)^2+\bigl(\alpha \omega (1-\frac{B_a\mu_s}{I\omega^2}-\frac{\mu_s}{I\gamma\omega})+\Gamma_g(1-\frac{\gamma(B_a+B_{an})}{\omega})\bigr)^2}},\label{eqn2a}
\end{eqnarray}

%\small
\begin{eqnarray}
\Omega_0&=&\frac{b_0\mu_s\sqrt{\alpha^2+(2-\frac{\gamma B_{an}}{\omega})^2}~~\exp{(i\theta_r)}}{I \sqrt{\bigl((B_a+B_{an})\gamma+\alpha\Gamma_g +\sigma \omega\bigr)^2+\bigl(\alpha \omega (1-\frac{B_a\mu_s}{I\omega^2}-\frac{\mu_s}{I\gamma\omega})+\Gamma_g(1-\frac{\gamma(B_a+B_{an})}{\omega})\bigr)^2}},\label{eqn2b}
\end{eqnarray}

where $\theta_m$ and $\theta_r$ are given by

% {\footnotesize
% %\small
% \begin{subequations}
% \begin{align}
% %%\mu_0&=&\frac{\gamma b_0\mu_s\sqrt{1+(\frac{\Gamma_g}{\omega}+\frac{\alpha \mu_s}{\gamma\omega I})^2}}{\sqrt{\bigl((B_a+B_{an})\gamma+\alpha\Gamma_g +\sigma \omega\bigr)^2+\bigl(\alpha \omega (1-\frac{B_a\mu_s}{I\omega^2}-\frac{\mu_s}{I\gamma\omega})+\Gamma_g(1-\frac{\gamma(B_a+B_{an})}{\omega})\bigr)^2}}~~\exp{(i\theta_m)},\label{eqn40} \\
% %%\Omega_0&=&\frac{b_0\mu_s\sqrt{\alpha^2+(2-\frac{\gamma B_{an}}{\omega})^2}}{I \sqrt{\bigl((B_a+B_{an})\gamma+\alpha\Gamma_g +\sigma \omega\bigr)^2+\bigl(\alpha \omega (1-\frac{B_a\mu_s}{I\omega^2}-\frac{\mu_s}{I\gamma\omega})+\Gamma_g(1-\frac{\gamma(B_a+B_{an})}{\omega})\bigr)^2}}~~\exp{(i\theta_r)},\label{eqn4a} \\
% %
% \theta_m&=&\tan^{-1}[\frac{\gamma \omega I}{\alpha \mu_s +\gamma \Gamma_g I}]-\tan^{-1}[\frac{\gamma \omega I(\alpha\Gamma_g+\gamma(B_a+B_{an})+\sigma\omega)}{\alpha \gamma B_a\gamma\mu_s+\Gamma_g I B_0\gamma^2+B_{an}\Gamma_g I\gamma^2+\alpha\omega\mu_s-\Gamma_g I\gamma\omega-\alpha I\gamma\omega^2}], \label{eqn3a}\\
% \theta_r&=&\tan^{-1}[\frac{2\omega-\gamma B_{an}}{\alpha\omega}]-\tan^{-1}[\frac{\gamma \omega I(\alpha\Gamma_g+\gamma(B_a+B_{an})+\sigma\omega)}{\alpha \gamma B_a\gamma\mu_s+\Gamma_g I B_0\gamma^2+B_{an}\Gamma_g I\gamma^2+\alpha\omega\mu_s-\Gamma_g I\gamma\omega-\alpha I\gamma\omega^2}].
% \label{eqn3b}  
% \end{align}
% \end{subequations}}

%\footnotesize
%\small
\begin{eqnarray}
\theta_m&=&\tan^{-1}[\frac{\gamma \omega I}{\alpha \mu_s +\gamma \Gamma_g I}]-\tan^{-1}[\frac{\gamma \omega I(\alpha\Gamma_g+\gamma(B_a+B_{an})+\sigma\omega)}{\alpha \gamma B_a\gamma\mu_s+\Gamma_g I B_0\gamma^2+B_{an}\Gamma_g I\gamma^2+\alpha\omega\mu_s-\Gamma_g I\gamma\omega-\alpha I\gamma\omega^2}], \label{eqn3a}
\end{eqnarray}

%\footnotesize
%\small
\begin{eqnarray}
\theta_r&=&\tan^{-1}[\frac{2\omega-\gamma B_{an}}{\alpha\omega}]-\tan^{-1}[\frac{\gamma \omega I(\alpha\Gamma_g+\gamma(B_a+B_{an})+\sigma\omega)}{\alpha \gamma B_a\gamma\mu_s+\Gamma_g I B_0\gamma^2+B_{an}\Gamma_g I\gamma^2+\alpha\omega\mu_s-\Gamma_g I\gamma\omega-\alpha I\gamma\omega^2}].
\label{eqn3b}  
\end{eqnarray}

% \begin{eqnarray}
%     \Omega_0&=&\frac{\gamma b_0\mu_s((i2+\alpha)\omega-i\gamma B_{an})}{\gamma B_a(\alpha \mu_s+\gamma I (\Gamma_g+i\omega))+\alpha \omega\mu_s+\gamma I (\Gamma_g+i\omega)(\gamma b_{an}+i(i+\alpha)\omega))}\\
%     % |\Omega_0|&=&\frac{b_0\mu_s\sqrt{\alpha^2+(2-\frac{\gamma B_{an}}{\omega})^2}}{I \sqrt{\bigl(\omega-(B_a+B_{an})\gamma-\alpha\Gamma_g\bigr)^2+\bigl(\alpha \omega (1-\frac{B_a\mu_s}{I\omega^2}-\frac{\mu_s}{I\gamma\omega})+\Gamma_g(1-\frac{\gamma(B_a+B_{an})}{\omega})\bigr)^2}}\\
%     % \arg{(\Omega_0)}&=&\tan^{-1}[\frac{2\omega -\gamma B_{an}}{\alpha\omega}]-\tan^{-1}[\frac{\gamma I(\alpha\Gamma_g+\gamma(B_a+B_{an})-\omega)\omega}{\alpha \gamma B_a\gamma\mu_s+\Gamma_g I B_0\gamma^2+B_{an}\Gamma_g I\gamma^2+\alpha\omega\mu_s-\Gamma_g I\gamma\omega-\alpha I\gamma\omega^2}]
% \end{eqnarray}

%% Such a resonance has a full-width at half-maximum linewidth of $\approx\sqrt{12\alpha^2\gamma^2 B_0^2}$.

\normalsize
Importantly, when $\sigma=-1$, $\mu_{0}$ and $\Omega_0$ reach maxima at $\omega_{FMR}=\gamma (B_a+B_{an})+\alpha\Gamma_g$. Typically, when a magnet is immobile, this resonance condition is met at $\omega_{FMR}=\omega=\gamma (B_a+B_{an})$ and is known as the ferromagnetic resonance (FMR). Mechanical motions of the levitated objects subjected to rotational damping change the frequency of FMR by $\alpha \Gamma_g$ with $\Gamma_g$ being proportional to $LP_g/M$, where $P_g$ is the residual gas pressure inside the levitation chamber \cite{KuhnStefan2017Odun}. When $\sigma=+1$, the rotating EM field opposes the precession of the magnetization \cite{DenisovPRL2006} and hence the induced magnetic moment in the $xy~$plane is several orders of magnitude smaller compared to the FMR case since, in our case, $\omega$ is very high (in hundreds of MHz to GHz range) and for most materials of interest $0<\alpha \ll 1$ \cite{Chang2014NYIG,MairFlaigPRB2017,BaratiPRB2017,KlinglerAPL2017}. Note that when $\mathbf{B_a}$ is applied along the $-z~$direction, a counterclockwise rotating ($\sigma=+1$) EM field can excite the FMR \cite{DenisovPRL2006}. Equations (\ref{eqn2a}) and (\ref{eqn2b}) also show that $\mu_0$ and $\Omega_0$ are proportional to the amplitude of the incoming electromagnetic field, as expected. Moreover, for a fixed $b_0$, when a levitated object gets bigger, i.e., $I\rightarrow \infty$, there is no mechanical precession ($\Omega_0\rightarrow 0$ and $\Gamma_g\rightarrow 0$), see (\ref{eqn2b}). In contrast, under the same condition, $\mu_0$ remains unchanged and can precess at $\omega$ as is the case when a magnet is attached to a large substrate and subjected to EM waves. For mesoscopic levitated objects on FMR (frequencies are in the hundreds of MHz to GHz range) and in high vacuum ($\Gamma_g \ll 1$, $\Gamma_g/\omega \ll 1$ and $\gamma B_{an}/\omega\ll 1$), we have 
%\footnotesize
% \begin{subequations}
% \begin{align}
% \mu_0&\approx \frac{b_0\gamma \mu_s}{\alpha \omega}~~\exp(i\theta_m),\label{eqn4a} \\
% \Omega_0&\approx\frac{2b_0\mu_s}{\alpha \omega I}~~\exp(i\theta_r).\label{eqn4b}
% \end{align}
% \end{subequations}

\begin{eqnarray}
    \mu_0&\approx \frac{b_0\gamma \mu_s}{\alpha \omega}~~\exp(i\theta_m),\label{eqn4a}
\end{eqnarray}

\begin{eqnarray}
\Omega_0&\approx\frac{2b_0\mu_s}{\alpha \omega I}~~\exp(i\theta_r).\label{eqn4b}
\end{eqnarray}
On FMR, the mechanical precession angle of a levitated rod and the corresponding displacement $\Delta x=\Delta y$ are
% \begin{subequations}
% \begin{align}
%     \theta_p&=\frac{|\Omega_0|}{\omega}=\frac{2b_0\mu_s}{\alpha  I\omega^2}. \label{eqn5a}\\
%     \Delta x&=L\sin\theta_p\approx \theta_p L.
%     \label{eqn5b}
% \end{align}
% \end{subequations}

\begin{eqnarray}
\theta_p&=\frac{|\Omega_0|}{\omega}=\frac{2b_0\mu_s}{\alpha  I\omega^2}. \label{eqn5a}
\end{eqnarray}
\begin{eqnarray}
\Delta x&=L\sin\theta_p\approx \theta_p L.
\label{eqn5b}  
\end{eqnarray}

From (\ref{eqn5a}) we observe that the precession angle, for a fixed $b_0$, decreases as $I$ increases. More importantly, $\theta_p$ decreases quadratically with $\omega$. In other words, a mechanical object cannot cope with the rising frequency of EM waves and becomes immobile when $\omega\rightarrow \infty$ for a finite $I$.

The detection of the precession frequency and thus the frequency of an incoming EM wave can be accomplished using the traditional optomechanical schemes which use laser interferometry and have been successfully deployed to detect rotations, precessions and translational motions \cite{RahmanRSI2018,AhnNatNano2020,ReimannPRL2018,RashidPRL2018,Kuhn17,GambhirPRA2015}. The precession frequency can also be measured using the rotational Doppler shift by illuminating the levitated objects using light with orbital angular momentum (OAM) \cite{CourtialPRL1998,Yao:11,LaveryScience2013}. The light with OAM can propagate parallel to the precession ($z$) axis. In this case, the frequency of the scattered or reflected light is shifted by multiples of the precession frequency $\omega$. In the discussion that follows, we take the rotational Doppler shift to be $\omega$. The change in the rotational Doppler shift due to an oblique incidence of the OAM beam onto the levitated particle is in 2nd order of $\theta_p$ ($\propto \theta_p^2$) and hence is negligible \cite{Qiu:19}.

\begin{figure}
%    \centering
    \includegraphics[width=8.4cm]{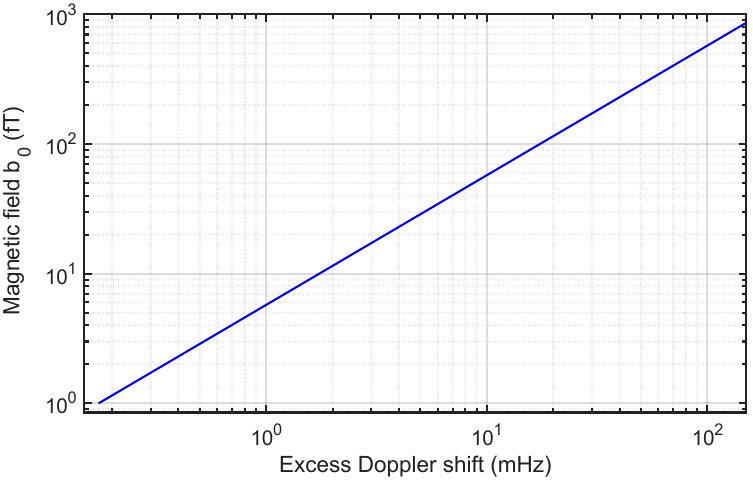}
    \caption{On ferromagnetic resonance, the amplitude of the magnetic field of an incoming EM wave as a function of the excess Doppler shift $(\Delta f-\omega r/\lambda_l)$ when the precession frequency or the frequency of the unknown EM wave is $\omega/2\pi=3~$GHz with a YIG of  $r=25~$nm and $L=1000~$nm as a sensor. In calculating $b_0$, we have used the saturation magnetization and Gilbert damping constant of YIG equal to $M_s=1.4\times 10^5~$A/m and $\alpha=1\times 10^{-4}$ \cite{KlinglerAPL2017}, respectively.}
    \label{fig2}
\end{figure}

The amplitude of the incoming EM waves can be measured using the linear Doppler shift. As the levitated rod precess about the $z~$axis, the linear velocity $\mathbf{v_r}$ of a point at the end of the rod located on the periphery is $\mathbf{v_r}=\omega (r+\Delta r)\hat{n}$, where $\mathbf{r}=r\cos{\omega t}~\hat{x}-r\sin{\omega t}~\hat{y}$, $\Delta r=\Delta x=\Delta y$, $\hat{n}=-\sin\omega t~\hat{x}-\cos{\omega t}~\hat{y}$ is the direction vector and $\mathbf{r}\perp \hat{n}$. The Doppler shift of a laser of wavelength $\lambda_l$ and propagating along the $y$ axis  is
\begin{eqnarray}
    %\Delta f &=&f_l\frac{\mathbf{v_r} . \mathbf{k_l}}{c}=\frac{\omega (r+\Delta x) \cos\omega t}{\lambda_l},
    \Delta f &=&f_l\frac{\mathbf{v_r} \cdot \mathbf{k_l}}{c}=\Bigl(\frac{\omega r }{\lambda_l}+\frac{\omega \Delta x }{\lambda_l}\Bigr)\cos\omega t,
    \label{eqn6}
\end{eqnarray}
%%%For the discussion that follows, we assume that a spatial filter placed at a location along the positive $+x~$axis is in use allowing the collection of scattered light shifted by $\omega (r+\Delta x)/\lambda_l$. which allow collecting scattered light over small solid angles can be used to reduce the linewidth further. Such a spatial filter can be placed at a location orthogonal to the propagation direction of the laser beam e.g., along the $+x~$axis. This would allow to collect scattered light shifted by only about $\omega (r+\Delta x)/\lambda_l$. 
%%In particular, high pass filters with cutoff frequencies $\omega (r+\Delta x)/\lambda_l$ can be used to study only the most Doppler shifted fraction of the scattered light.
where $f_l=\frac{c}{\lambda_l}$ is the frequency of the linearly polarized laser in vacuum, $c$ is the speed of light in vacuum and $\mathbf{k_l}=\hat{y}$ is a unit vector parallel to the laser propagation direction. The first term in (\ref{eqn6}) represents the Doppler shift when the rod spins about its long axis, while the second term is an excess shift that a precessing rod produces, and we use it to determine the amplitude ($b_0$) of the incoming EM wave (see Figure \ref{fig2}). The total Doppler shift $\Delta f$ can be measured by illuminating the precessing rod using a linearly polarized laser and collecting the scattered light. Subsequently, the scattered light can be mixed with a reference beam or a local oscillator (LO), which has not interacted with the rod. Mixing of the scattered and reference beams creates sidebands at $f_l\pm \Delta f$. The linewidths of the sidebands, however, can be broad, i.e., $\omega (r+\Delta x)/\lambda_l$ due to the oscillatory term in (\ref{eqn6}). Specifically, when $\mathbf{k_l}\perp \hat{n}$ ($\omega t=\pi/2,~3\pi/2,\ldots$) and $\mathbf{k_l}\parallel \hat{n}$ ($\omega t=0,~\pi,\ldots$), the Doppler shifts are zero and $\pm \omega (r+\Delta x)/\lambda_l$, respectively. Other angles between $\hat{n}$ and $\mathbf{k_l}$ create differing amounts of shifts, i.e., $0<|\Delta f| < \omega (r+\Delta x)/\lambda_l$. When mixed with the unperturbed laser beam, these variable amounts of Doppler shifts create broad sidebands. However, to reduce the width of the Doppler shift, only one end of the levitated rod can be illuminated. Moreover, pinholes, a form of spatial filter, placed at a location orthogonal to the propagation direction of the laser beam, e.g., along the $+x~$axis, can be used to collect light scattered over a small solid angle and shifted by only about $+\omega (r+\Delta x)/\lambda_l$. A side effect of such spatial filtering might be to reduce the amount of scattered light collected, which in turn can degrade the signal. This can be mitigated using a strong local oscillator. A strong LO can also minimize the effect of laser shot noise. To reduce the width of the sidebands further, once the signal is in the electrical domain, electrical filters can be used to selectively look at certain frequencies. To aid electrical filtering, the frequency of the LO can be adjusted such that the scattered light of frequency $[c/\lambda_l+\omega (r+\Delta x)/\lambda_l]$, after mixing with the LO, appears close to DC. This would allow using narrow, e.g., mHz, bandwidth electrical band-pass filters. With spatial and electrical filters in place, it is viable that scattered light shifted in frequency by $\omega (r+\Delta x)/\lambda_l$ can be detected. Taking $\cos\omega t\approx 1$ and by substituting $\Delta x$ from (\ref{eqn5b}) in (\ref{eqn6}), we find
% From the measured linear Doppler shift $\Delta f$, $b_0$ can be retrieved. 
\begin{eqnarray}
% \nonumber
%     \Delta f &=&\frac{\omega (r+\Delta x)\cos\omega t}{\lambda_l}=\frac{\omega r}{\lambda_l}+\frac{\omega \Delta x}{\lambda_l}\\
%     \nonumber
%            \Delta x &=& \frac{\lambda_l\Delta f}{\omega}-r\\
%            \nonumber
%            \theta_p&=&\frac{\lambda_l\Delta f}{\omega L}-\frac{r}{L}\\
%            \nonumber
%            \frac{2b_0\mu_s}{\alpha  I\omega_{FMR}^2}&=&\frac{\lambda_l\Delta f}{\omega L}-\frac{r}{L}\\
%            \nonumber
%            b_0&=&\frac{\alpha  I\omega^2}{2\mu_sL}\Bigl(\frac{\lambda_l\Delta f}{\omega}-r\Bigr)\\
%            b_0&=&\frac{\alpha  I\omega }{2\mu_sL}\Bigl(\Delta f \lambda_l-\omega r\Bigr).\\
   %                   b_0&=&\frac{\alpha  I}{2\mu_sL}\Bigl(\frac{\Delta f \omega \lambda_l}{\cos\omega t}-r\omega^2 \Bigr).
                      b_0&=&\frac{\alpha  I}{2\mu_sL}\Bigl(\Delta f \omega \lambda_l-r\omega^2 \Bigr).
    \label{eqn7}
\end{eqnarray}

All parameters in (\ref{eqn7}) are either known or can be measured during the experiment. For example, $\omega$ and $\Delta f$ can be found from the rotational and linear Doppler shifts described before, while $\lambda_l$ is a known laser wavelength. The prefactor $\frac{\alpha  I}{2\mu_sL}$ in  (\ref{eqn7}) and the radius $r$ of the levitated rod can be found in-situ by calibrating the sensor using known $b_0$, $\omega$ and $\Delta f$. The length and radius of a levitated rod can also be estimated from electron microscopy performed on the rods prior to levitation. The Gilbert damping constant can be found by plotting the mechanical precession frequency $\omega_{FMR}=\gamma(B_a+B_{an})+\alpha\Gamma_g$ as a function of residual gas pressure or $\Gamma_g$. The slope of such a graph is $\alpha$.

%From (\ref{eqn7}), it can be observed that objects of small moment of inertia and relatively long in size ($L$) are better for sensing. 

%To be used as sensors, rods of any magnetic material can be used as long as they can be levitated. 

\section{Results and Discussion}\label{sec3}
As evident in (\ref{eqn7}), strong magnets with large $\mu_s$, such as samarium-cobalt, are preferable as sensors. However, such magnets have large Gilbert damping constants (on the order of $10^{-2}$), which degrade the sensor performance. For the discussion that follows, we take yttrium iron garnet (YIG) as a model material \cite{CHEREPANOV199381,Serga_2010}, which is a weak magnet but has the lowest known Gilbert damping constant $\alpha\approx 2.7\times 10^{-5}$ \cite{KlinglerAPL2017} (the ratio $\alpha/\mu_s$ is superior for YIG compared to other magnets). Moreover, YIG is an electrical insulator, implying it is not susceptible to eddy current and the associated heating. YIG rods or cylinders can be fabricated starting from high-quality bulk YIG single crystal using micro-fabrication \cite{KuhnNano2015,HAUSMANN2010621, Cheung_2006,SHI2023107311}. As fabricated rods can be separated from substrates by lasers and launched directly to traps~\cite{KuhnNano2015}. Rods can also be separated from substrates by mechanical pulverization and collected for putting into traps using a nebulizer \cite{seberson2020YIG}, a piezoelectric shaker \cite{GambhirPRA2015} or laser ablation \cite{DaniaPRL2024}. Levitation can be carried out using a Paul trap, which uses oscillating electric fields (frequency in kHz) and can levitate any object as long as they are charged  \cite{PennyPRR2023,DelordNJP2017,Bykov:23,jin2023quantum}. Paul traps have been used for demonstrating ultra-stable rotational and translational motions of levitated objects with linewidths less than a mHz \cite{jin2023quantum} and a $\mu$Hz \cite{DaniaPRL2024}, respectively.  Due to the small amount of light used in a Paul trap for the detection of levitated particles, resistive torques associated with the radiation pressure and light scattering are negligible~\cite{RahmanPRA2023,Jain2016}. The sensor can be configured in a fixed or a tunable frequency mode. In the fixed frequency mode, the externally applied bias magnetic field remains constant, and an incoming EM wave of frequency $\gamma (B_a+B_{an})$ initiates precession. In contrast, in the tunable mode, $B_a$ is adjusted until a resonance in mechanical precession is detected. Note that the frequency of the targeted EM waves (MHz to GHz, see below for further details) and thus the mechanical precession frequency of the levitated rod is substantially different from the translational/trap frequencies ($\approx$kHz) of the levitated rods \cite{PennyPRR2023,DelordNJP2017,Bykov:23,jin2023quantum}, and hence can be easily distinguished.

The smallest field amplitude $b_0$ that the sensor is sensitive to is dictated by the experimental conditions portrayed in (\ref{eqn7}). We assume that all the physical parameters of the levitated rod, i.e., $\mu_s$, $\alpha$, $L$, $r$ and $I$ or their combinations are known through sensor calibrations and other measurements such as electron microscopy. We also assume that the frequency of the unknown EM wave has already been determined through the measurement of the rotational Doppler shift or optomechanical measurements described before. The measurement accuracy of $b_0$ depends on how accurately $\omega$ and $\Delta f$ can be measured. Fortunately, the measurement precision of frequencies is excellent, and fractional accuracies of $10^{-18}$ or better using optical clocks have been achieved \cite{Fortier:26}. When converted to the microwave domain, this amounts to an accuracy of micro hertz or better. Figure \ref{fig2} shows the amplitudes of EM waves for a varying amount of excess Doppler shifts ($\Delta f-\omega r/\lambda_l$) with a YIG rod of size $r=25~$nm and $L=1000~$nm. For a $1~$mHz excess Doppler shift, the weakest detectable magnetic field is $\approx$6~fT or $1.8~\mu$V/m when converted to electric fields. In principle, even weaker fields can be detected since rotation frequencies with accuracies better than a mHz have been measured using Paul \cite{jin2023quantum} and optical traps \cite{KuhnStefan2017Odun}. Nevertheless, the minimum detectable field is ultimately determined when the stochastic torque due to gas molecules $\sqrt{4 k_bT I\Gamma_g/t_m}$ \cite{AhnNatNano2020,ZhujingPRA2017}, where $k_b$ is the Boltzmann constant, $T$ is the temperature, and $t_m$ is the measurement time. Taking the corresponding thermal magnetic field to be $b_{th}$, the resulting torque is $\boldsymbol{\mu}\times \mathbf{b_{th}}$. In other words, we have
\begin{eqnarray}
    b_{th}=\frac{1}{\mu}\sqrt{\frac{4 k_bT I\Gamma_g}{t_m}},
    \label{eqn8}
\end{eqnarray}
where we have set $\boldsymbol{\mu}\perp \mathbf{b_s}$. For a YIG rod levitated in UHV (gas pressure $P_g=10^{-10}~$mBar) with $t_m=10~$s and $\Gamma_g\approx (2rLP_g/M)\sqrt{2\pi m_g/k_bT}$ \cite{KuhnStefan2017Odun}, we have $b_{th}=5\times 10^{-15}~$T, where $m_g$ is the mass of the gas molecules. So, the thermal limit of the smallest detectable $b_0$ is about the same as that found using a $1~$mHz Doppler shift.

The ability to tune frequencies continuously over a large frequency range is an essential attribute of a versatile EM sensor. In our case, this can be achieved by adjusting $B_a$, which changes the FMR frequencies, see (\ref{eqn2a}) and (\ref{eqn2b}). $B_a$ can be delivered using an electromagnet or a permanent magnet. The strength of $B_a$ can be adjusted by changing the current passing through an electromagnet or the distance of the permanent magnet from the levitated object. In either case, as $B_a$ changes, the frequency of the ferromagnetic resonance changes, allowing the detection of unknown frequencies in a continuous manner. The highest frequency that our sensor can detect is limited by the maximum tensile stress ($\approx$$\rho_m \omega^2r^2$), arising from the mechanical precession, that a levitated magnet can withstand \cite{SchuckScience2018,AhnPRL2018}. Mechanical rotations over $5~$GHz have been demonstrated \cite{AhnNatNano2020} and tens of GHz have been predicted \cite{AhnPRL2018}. The lowest frequency that a precessing magnetometer can sense is determined by the saturation magnetization and/or the magnetocrystalline anisotropy of a levitated magnet. In this context, weak magnets such as YIG are preferable since their saturation magnetization is low. Using levitated YIG rods as sensors, frequencies down to $200~$MHz can be detected~\cite{LeeJAP2016,BryantAPL1992}. Another key aspect of our sensor is that it can be switched off when required \cite{FancherIEE2021}, important in defense applications, by changing the strength of the DC bias field. This is in contrast with the classical antennas.

The selectivity of a magnetometer is defined as its ability to detect signals around a given center frequency. It is also known as the Q-factor of a sensor \cite{FancherIEE2021}. At a given $B_a$, our magnetometer is sensitive to frequencies within the ferromagnetic resonance linewidth~$\approx\sqrt{12\alpha^2\gamma^2 B_a^2}$, which is dictated by the Gilbert damping constant. For high selectivity, materials with low $\alpha$'s are required. Here, YIG is an excellent candidate, which is known for its extremely narrow FMR linewidth \cite{MairFlaigPRB2017,Chang2014NYIG,Spencer1961}. Using YIG rods as sensors, frequency selectivity of less than a MHz can be achieved. Importantly, within this frequency band, our magnetometer can resolve frequencies with a sub-hertz resolution. This is a result of it being possible to measure mechanical precession frequencies with resolutions better than a mHz \cite{KuhnStefan2017Odun,jin2023quantum}.

%%In addition to broad tunability, the ability to differentiate different frequencies or the so-called frequency resolution is an important attribute of an EM sensor. In our case, this can be mHz or lower. 

%%In essence, a precessing levitated magnet can detect frequencies from MHz upto tens of GHz. 

One of the key attributes of a versatile sensor is its ability to remain sensitive when the amplitudes of signals vary widely. This is known as the dynamic range of a sensor. In our case, the angle of precession (Figure \ref{fig1}) of the levitated magnet is determined by the strength of the unknown oscillating magnetic field, see (\ref{eqn5a}). Assuming the magnet is already precessing at $\omega$, as the strength of the EM field increases, the angle of precession of the magnet increases as well. This remains true as long as the strength of the unknown magnetic field is $\ll$$B_a$. If the unknown field becomes comparable to $B_a$, the approximation made in deriving (\ref{eqn2a}) and (\ref{eqn2b}), e.g., $b_0\ll B_a$, breaks down. Importantly, as the precession angle increases with the increasing $b_0$, the detection of precession becomes easier, implying an enhanced sensitivity. This is in contrast with other magnetometers, which become less sensitive as the strength of the field increases \cite{Graham2023, Couedo2019}. Generally, $B_a$ is in tens of millitesla, implying a large dynamic range, e.g., femtotesla to millitesla for a levitated magnet-based magnetometer.

The direction of arrival (DA) of an EM wave is important in many areas of engineering and physics, including in defense \cite{RobinsonAPL2021,HuangZheng2022} and astrophysics \cite{Thornton2013}. In our case, the direction of arrival $\mathbf{k}$ can be found from the direction of $\mathbf{B_a}$. In deriving (\ref{eqn2b}), we assumed that $\mathbf{b_s}$ is in the $xy~$plane and hence, in this simple case, the direction of arrival is along the $z~$axis. To further differentiate between the arrival along the $-z$ and the $+z$ directions, the sensor, due to its small size (equivalent to a small vacuum chamber \cite{Arita2013}), can be temporarily blocked using an EM absorber \cite{ELMAHAISHI20222188} from one of the two sides. In the event the EM arrives from the blocked side, the levitated magnet will stop precessing, thus determining the arrival direction. For an arbitrary arrival, the direction of $\mathbf{B_a}$ can be adjusted such that $\mathbf{b_s}$ becomes perpendicular to $\mathbf{B_a}$. This can be achieved by mounting the bias magnet on a rotary mount. In the case when $\mathbf{k}\perp \mathbf{B_a}$, there is no torque and the magnet cannot precess.

Finally, it is instructive to consider potential sources of noise that can degrade the performance of our sensor. One such source is the fluctuation in the strength of the DC magnetic field $B_a$. Such fluctuations would change the FMR frequencies and thus the driving torque (see (\ref{eqn2a}) and (\ref{eqn2b}) and the discussion surrounding it). In turn, this can cause uncertainties in the minimum detectable fields. The fractional fluctuation of the bias magnetic field, when delivered using a samarium cobalt permanent magnet, is better than one part per $10^{13}$ with a magnet of volume $10\times 10\times 10~$cm$^3$ (field fluctuation is inversely proportional to the square root of the volume of the magnet) \cite{ErnstJAP1984}. For $B_a=0.10~$T ($\omega/2\pi\approx 3~$GHz), this is equivalent to a frequency fluctuation of $2\times 10^{-4}~$Hz. Charge or the drive voltage fluctuations in a Paul trap when such a trap is used for levitation may constitute a source of noise. However, using such a trap, highly stable rotational motion of levitated diamonds (linewidth $<10^{-4}~$Hz) has been demonstrated \cite{jin2023quantum}. Paul traps have also been used to demonstrate linewidths in the nano hertz regime using levitated silica nanoparticles in UHV \cite{DaniaPRL2024}. Consequently, a Paul trap is not a limiting factor for our sensor. Microwaves generated by a blackbody (BB) source are another potential source of noise. Such sources are omnidirectional and incoherent. Nevertheless, for our calculations we consider such a source produces plane waves and find the amplitude of the relevant magnetic fields to be $b_{bb}(\omega)=\sqrt{\frac{\mu_0 \hbar \omega^3 \Delta \omega}{2\pi^2c^3[\exp(\hbar\omega/k_BT)-1]}}$ \cite{Mansuripur2017}, where $\hbar$ is the reduced Planck constant and $\Delta \omega$ is the bandwidth of the BB source around $\omega$. In the frequency band of our interest (hundreds of MHz to a few tens of GHz) and with $\Delta\omega/2\pi=1~$Hz, we find $b_{bb}<4\times 10^{-16}~$T, which is negligible. Traditional $1/f$, as well as laser intensity noise when a relatively good laser is in use, are not significant in the frequency range of our interest.

%%As a result, our sensor is robust against main sources of noise e.g., charge/voltage and bias field fluctuations. 

%%However, highly stable, better than a part-per-billion fluctuation, electromagnets are available \cite{VanDyckRSI1999,BrittonPRA2016}. Likewise, 

%%hen converted to an electric field, this is equivalent to $3\times 10^{-7}~$V/m.  

\section{Conclusions}
In conclusion, we have theoretically shown that a levitated magnet in a vacuum is capable of detecting weak electromagnetic fields of femtotesla level. If achieved experimentally, this would be better than the performance of existing sensors \cite{GordonAIPAdv2019,JingMingyong2020Asrb,FancherIEE2021}, with the added benefit of the simplicity of our proposed experiment. The overall sensor size, including a vacuum chamber \cite{Arita2013} and a Paul trap \cite{jin2023quantum} can be $<$1~cm$^3$. We have also shown that the new magnetometer can be continuously tuned between hundreds of MHz and tens of GHz and remains sensitive when the field strength varies between femtotesla to millitesla. Within the frequency range of operation, the new magnetometer can measure frequencies with a resolution better than a mHz and a selectivity of about a MHz. We envisage that due to its ability to detect femtotesla level field and be configured for sensing different frequencies by merely changing the externally applied magnetic field, the new magnetometer can be useful in fields such as biomedicine \cite{LiMW2013,Zhao2022}, search and rescue \cite{NguyenMW2019} and defense \cite{NguyenMW2019} where the ability to sense at different frequencies is crucial.

%%\bibliography{Bibliography}


\end{document}